# China RealDID: Verifiable Credentials Anchored in Legal Identity

**Yifan He**
Founder and Chief Executive Officer, Red Date Technology (Hong Kong) Ltd.

yifan.he@reddatetech.com

ORCID: 0009-0005-1048-180X

---

> **Working Paper** — Comments and circulation welcome. This is a working paper circulated for discussion and feedback. A condensed version is planned for journal submission. Operational metrics, including DID issuance volume, consortium composition, and end-to-end credential issuance and verification latency, will be added in a future revision once proper authorization is received.
>
> **Author's note.** The author is CEO of Red Date Technology (Hong Kong) Ltd., co-developer of the RealDID system described in this paper. Architectural descriptions reflect insider knowledge; normative claims should be read with this affiliation in mind. A formal Competing Interests declaration appears before the References.

## Abstract

Verifiable credentials (VCs) and decentralized identifiers (DIDs) enable selective disclosure without centralized identity providers, but suffer a structural weakness: without a trusted identity root, verifiers cannot distinguish a genuine credential holder from a fabricated identity. National legal identity systems provide biometric-grounded verification but impose three costs: every verifier must collect the subject's full personally identifiable information, verification infrastructure concentrates on a single state-operated API, and the state observes every verification transaction. We present China's RealDID as a case study bridging these paradigms through a three-layer architecture—CTID (centralized legal identity), RealDID (decentralized identifier anchor on an open permissioned blockchain), and VCs with selective disclosure—evaluated against an adversary model of five adversary classes and six security goals. The central mechanism is a content-blind government relay: the state authenticates VC participants and counter-signs every credential but cannot read the payload, which the issuer encrypts directly to the holder's public key. We describe the VC lifecycle, triple-signature chain, template registry with open issuer participation, selective-disclosure mechanism, and the architecture's metadata-level privacy limits—including the credential graph accumulated at the relay and the presentation-linkability gap from single-DID reuse. The design yields an asymmetric, state-bounded trust model: the state cannot impersonate users or read credential contents, while the user cannot fabricate legal identity or evade state-side metadata observation. We analyze regulatory alignment with China's Personal Information Protection Law and the EU's GDPR, including the tension between immutable on-chain registries and erasure rights, and discuss generalizability through cross-border deployments with Singapore and Hong Kong.

**Keywords:** RealDID • verifiable credentials • decentralized identity • CTID • selective disclosure • digital identity • privacy • authentication • China

## 1. Introduction

Verifiable credentials (VCs) and decentralized identifiers (DIDs), standardized by the W3C [1, 2], enable cryptographic selective disclosure: a holder can present a subset of claims from an issuer-signed credential to a verifier, who validates the issuer's and holder's signatures without querying a central authority. This model addresses several long-standing privacy weaknesses of centralized identity verification — verifier-side PII over-collection, single-API cost concentration, and state visibility into all verification transactions. However, the decentralized identity ecosystem carries a complementary security gap: it lacks *legal anchoring*. A self-issued DID provides no guarantee that the entity behind it corresponds to a real, legally-recognized person. In the absence of a trusted identity root, a verifier cannot distinguish between a genuine credential holder and a fabricated identity backed by a self-generated key pair. Cryptographic verification of a credential's signatures confirms *who issued it* and *who presents it*, but says nothing about whether the underlying identity is real — a property that only a state-operated, biometric-grounded legal identity system can provide.

This tension — between the privacy and decentralization benefits of VCs and the legal-anchoring requirement that only a centralized state root can satisfy — is the problem this paper addresses. We present China's RealDID system as a case study that bridges centralized legal identity with decentralized verifiable credentials, and we evaluate the result against an explicit security model.

RealDID is a decentralized identity layer built on top of CTID (Cyber Trusted Identity), China's centralized identity verification infrastructure operated by the responsible national authority. After a user invokes CTID's per-session biometric verification — matching a live facial image against their pre-existing CTID record, which is populated automatically when a citizen obtains their national ID card — the responsible state authority issues a `did:sic:` identifier (State Information Center) on an open permissioned blockchain governed by a multi-stakeholder consortium of public and private sector entities. The user generates their own cryptographic key pair locally; the government records the mapping between the RealDID and the CTID identity in an internal, off-chain linkage table but never publishes it on-chain. The government has no access to the user's private key and therefore cannot sign as the user, present credentials, or impersonate the DID subject. Verifiable credentials are issued against RealDIDs through a government-mediated but **content-blind relay**: the government verifies that both the issuer and the holder are authentic, but cannot view the credential contents, which the issuer encrypts directly to the holder's public key before the credential reaches the relay. The government never holds a key capable of decrypting the payload. Every VC presentation carries a **triple-signature chain**: the issuer signature attests to the credential's authority, the government counter-signature attests to the authenticity of both parties, and the holder signature on each presentation proves current control of the DID private key and prevents replay.

This architecture yields what we call an **asymmetric, state-bounded trust model**: the state retains legal anchoring, deanonymization, revocation, and per-credential metadata logging, yet is cryptographically prevented from impersonating users or reading credential contents; the user controls selective disclosure but cannot fabricate a legal identity or evade state-side metadata observation. The user is bounded by cryptography; the state is bounded by cryptography plus legal-process constraints on capabilities it both holds and operates. We frame this as **decentralized verification with centralized legal anchoring** rather than fully decentralized identity — a more honest baseline than stronger "mutual constraint" rhetoric, developed in detail in §4.4 and §6.2.

We evaluate the architecture against an explicit adversary model (§3.0) comprising five adversary classes: malicious verifier (A1), semi-honest issuer (A2), semi-honest government relay (A3), network observer (A4), and consortium-majority adversary (A5). We define six security goals — issuer authentication (G1), holder authentication (G2), selective disclosure of contents (G3), content blindness at the relay (G4), verifier privacy (G5), and accountability under legal process (G6) — and map each to the architectural mechanism that realizes it, the adversary classes it addresses, and the assumptions and limitations that bound the claim.

This paper makes three contributions:

1. **Security architecture**: We describe the RealDID three-layer architecture (CTID → RealDID → VCs) in protocol-level detail, including the DID issuance flow, key custody model, VC template registry with open issuer participation, full VC lifecycle, triple-signature chain, content-blind relay design, and the selective-disclosure mechanism. We provide a structured security-claim table mapping each of the six security goals to its architectural mechanism, adversary coverage, and bounding assumptions. We name the architecture's security and privacy limits explicitly: the credential-graph metadata that the relay accumulates, the presentation-linkability gap from single-DID reuse, and the reliance on consortium consensus rather than cryptographic enforcement against a majority adversary.

2. **Policy and regulatory analysis**: We analyze the architecture's alignment with China's Personal Information Protection Law (PIPL) and compare it with the EU's General Data Protection Regulation (GDPR) and eIDAS 2.0, including the tension between immutable on-chain registries and data-subject erasure rights. We characterize the dual-root trust model and its asymmetry, and identify verifier privacy as an under-appreciated dimension of identity infrastructure — protected by decentralized verification but structurally exposed by centralized API models.

3. **Generalizability**: We argue that the three-layer model is chain-agnostic and applies to any jurisdiction with a centralized legal identity system. We discuss this through in-progress cross-border deployments with Singapore and Hong Kong; empirical evaluation is reserved for future work.

The remainder of this paper is organized as follows. Section 2 provides background on CTID, RealDID, W3C Verifiable Credentials, and related work. Section 3 presents the adversary model and security goals, followed by a gap analysis through five problem scenarios. Section 4 details the RealDID identity-anchor architecture. Section 5 describes the VC ecosystem built on RealDID. Section 6 provides policy and governance analysis. Section 7 discusses generalizability to other jurisdictions. Section 8 concludes.

# 2. Background

## 2.1 CTID: China's Centralized Identity Infrastructure

The Cyber Trusted Identity (CTID) system is China's national online identity authentication platform, operated by the responsible national authority. CTID anchors legal identity to biometric and documentary evidence collected through the national ID-card issuance process: when a Chinese citizen applies for a national ID card, the issuing authority records the citizen's biometric (facial image) and documentary data, and CTID is populated automatically from that record. No separate user-initiated CTID registration step is required. Once verified, CTID binds the user's legal identity — name, identification number, photograph, and associated demographic data — to an authenticated session or token.

CTID exposes a single permission-based verification API gateway. Authorized service providers — primarily financial institutions, telecommunications carriers, and government platforms — submit an authentication request containing the subject's claimed name, identification number, and facial image, and CTID returns only a yes/no confirmation of whether the submitted data matches its records. The API is not publicly accessible; access requires formal authorization and is limited to a relatively small set of entities. This architecture concentrates three burdens on the state: (1) full infrastructure cost for national-scale verification traffic, (2) sole responsibility for access control decisions, and (3) complete visibility into every verification transaction — who verifies whom, when, and in what context.

This API-gateway form is the original CTID deployment model. CTID is currently being upgraded toward a complementary app-based mode of operation — structurally analogous to the national identity applications deployed in Singapore and Hong Kong — in which the user authenticates through a state-issued mobile application rather than (or in addition to) the verifier-submitted query model described above. The architectural analysis in the remainder of this paper applies to both modes, since the RealDID layer is decoupled from how CTID itself is exposed to verifiers.

## 2.2 RealDID: The Decentralized Identity Layer

RealDID is a decentralized identity system built as an extension to CTID. Its core premise is to decouple identity *verification* from identity *authentication*: CTID remains the root of legal identity verification (biometric + documentary), but RealDID shifts ongoing authentication and credential exchange to a decentralized infrastructure.

RealDID operates on an **open permissioned blockchain** — specifically **Guoxin Chain** (formerly Yan'an Chain [31]), built on the FISCO BCOS open-source blockchain framework [33] and operated by China's Blockchain-based Service Network (BSN), launched 9 June 2023 — governed by a multi-stakeholder consortium of government agencies and private sector entities. The chain is permissioned at the validator level — only consortium members may operate nodes and write blocks — but publicly readable: any party with network access can query DIDs, resolve DID documents, and verify cryptographic signatures without authorization. This design is driven by three constraints. First, public permissionless blockchains require cryptocurrency gas fees, which are legally restricted in China. Second, a private single-entity chain would defeat the purpose of decentralized verification by placing one party in unilateral control. Third, the consortium model distributes governance across multiple stakeholders, ensuring that no single entity — including the state — can unilaterally alter the DID registry, template registry, or revocation registry.

The RealDID issuance protocol proceeds as follows:

1. **CTID verification**: The user authenticates through CTID's biometric verification (facial recognition and ID document matching).
2. **Key generation**: The user generates a cryptographic key pair locally. The public key is submitted to the government; the private key never leaves the user's device.
3. **DID issuance**: The government issues a `did:sic:<address>` to the permissioned chain, associating the DID with the user's public key.

4. **Linkage recording**: The government internally records the mapping between the CTID identity and the `did:sic:` — this linkage table is held exclusively by the government and is never published on-chain.
5. **Key custody**: The user retains exclusive control of the private key. The government cannot sign transactions, present credentials, or otherwise act as the DID subject.

## 2.3 W3C Verifiable Credentials

The W3C Verifiable Credentials Data Model [2] defines a standard for cryptographically verifiable digital credentials. A VC is a set of claims about a subject, issued by an authoritative issuer, and held by the subject in a digital wallet. The holder can present the VC — or a subset of its claims — to a verifier, who cryptographically validates the issuer's signature and the holder's control of the credential.

The W3C Decentralized Identifiers specification [1] defines DIDs as globally unique, resolvable identifiers that do not require a centralized registration authority. A DID resolves to a DID document containing the subject's public keys and service endpoints. DIDs are typically anchored to a distributed ledger or similar verifiable data registry, enabling anyone to resolve them without querying a central authority.

Three properties of the VC model are central to the RealDID architecture. **Selective disclosure** allows a holder to reveal only a subset of claims from a VC — for example, proving age over 18 without revealing date of birth. **Issuer authentication** ensures that every VC carries a cryptographic signature from its issuer, which verifiers can validate against the issuer's DID document. **Presentation authentication** ensures that the party presenting a VC is its legitimate holder, via a holder signature bound to the DID's private key.

## 2.4 Related Work

Several lines of work are relevant: (i) self-sovereign and decentralized identity frameworks, (ii) state-led mobile and verifiable credential deployments, (iii) bridges between national identity infrastructure and decentralized credentials, and (iv) the underlying cryptographic primitives for selective disclosure.

**Self-sovereign and decentralized identity frameworks.** Allen's "Path to Self-Sovereign Identity" [19] articulates the ten principles (existence, control, access, transparency, persistence, portability, interoperability, consent, minimization, protection) that the SSI ecosystem has built on. Civic identity on blockchain has been explored by Sovrin [5] (a purpose-built identity ledger built on Hyperledger Indy, which also originated the AnonCreds anonymous-credential scheme [22]) and uPort [32] (Ethereum-based self-sovereign identity). These systems emphasize user sovereignty but assume self-asserted or socially-attested identity roots, limiting their legal standing for applications requiring government-recognized identity. The Trust over IP Foundation has developed a four-layer governance model [23] that complements the technical W3C stack with governance and assurance layers; RealDID's consortium-governed registries occupy that layer in a way that the W3C documents leave deliberately unspecified.

**State-led mobile and verifiable credential deployments.** The European Union's eIDAS 2.0 regulation [3] mandates the European Digital Identity Wallet (EUDI Wallet), which enables citizens to store and present digital credentials with selective disclosure. The accompanying EUDI ARF [15] specifies SD-JWT [12] and ISO/IEC 18013-5 mobile driver's licenses (mDL) [16] as the interoperable credential formats; together with Hyperledger Indy/AnonCreds [22], these are the principal production-grade reference designs against which RealDID can be compared. The mDL standard in particular has reached real-world deployment in multiple jurisdictions [25]. NIST SP 800-63 [26] provides the U.S. assurance-level baseline for digital identity, including remote identity proofing at the highest assurance level (IAL3), which like CTID requires biometric verification. The trust model in eIDAS 2.0 is federated: member states authenticate identity through their national systems and issue credentials into the wallet. The EUDI Wallet achieves user-controlled selective disclosure through a federation of national trust roots rather than a single state root.

**Bridges between national identity and decentralized credentials.** India's Aadhaar system has moved from being a candidate identity root for verifiable credentials to an actual one: the Aadhaar (Authentication and Offline Verification) Amendment Regulations, 2024 [4] formalize the *Aadhaar Verifiable Credential* (AVC) as a recognized offline verification mode, allowing holders to share a UIDAI-signed credential containing selected demographic fields (and the last four digits of the Aadhaar number) rather than the full Aadhaar identity. Architecturally, AVC is closer to RealDID's selective-disclosure ethos than the original Aadhaar API model is — but AVC remains *issuer-rooted* in UIDAI without an intermediate decentralized identifier layer, and Aadhaar's biometric-centralized architecture has been the subject of a substantial critical literature documenting exclusion harms and centralized-biometric-storage risks [20, 21]. The presence of AVC in production makes India the most directly relevant comparator for the RealDID model: same problem class (state legal identity bridging to user-held credentials), distinct architectural choice (no on-chain DID layer; UIDAI as sole anchor). Estonia's e-Residency and X-Road infrastructure [27] is yet another model — a federated set of state-operated services around a national PKI — and is a useful reference for how mature state digital identity can interoperate with credential issuance without a ledger.

**Cryptographic primitives for selective disclosure.** Cryptographic selective disclosure schemes — BBS+ signatures [10], Camenisch-Lysyanskaya signatures [11], and the IETF SD-JWT specification [12] — allow a holder to prove possession of an issuer-signed credential while revealing only a subset of its claims, without the holder having to re-sign a redacted version. Anonymous-credential schemes (IDEMIX [24], AnonCreds [22]) additionally provide presentation unlinkability, so that two presentations from the same holder cannot be linked to a common identity by colluding verifiers. RealDID uses SD-JWT [12] for cryptographic selective disclosure (§5.3), providing the same hash-commitment mechanism as the EUDI credential format; the unlinkability gap from single-DID reuse is discussed in §6.4.

**RealDID's distinguishing features.** Relative to this prior work, the features that distinguish RealDID are: (1) **content-blind government relay** — the state authenticates VC participants without viewing credential payloads; (2) **user-held private keys with state-exclusive linkage** — the state can deanonymize a DID under legal process but cannot impersonate; (3) **verifier privacy at the application layer** — on-chain queries do not require authentication to a central authority, in contrast to permissioned verification APIs; and (4) **chain-agnostic governance** — the architecture prescribes chain *properties* (public readability, multi-stakeholder governance of registries) rather than a specific chain type. None of these is individually novel as a cryptographic technique; the contribution is in their integration with a centralized legal identity root, in a regulatory environment that constrains the chain layer, and in the explicit naming of the asymmetries that follow.

## 3. Privacy and Verification Gap Analysis

Before presenting the RealDID architecture in detail, we establish the problems it addresses. This section first specifies the adversary model and security goals against which both centralized CTID-style verification and the proposed RealDID architecture are evaluated (§3.0), then analyzes five structural gaps in centralized identity verification, each illustrated with a concrete scenario (§3.1–§3.5).

### 3.0 Adversary Model and Security Goals

We consider five adversary classes. The model is informal — a fully game-based treatment is out of scope for a case-study paper — but it pins down which capabilities each adversary holds and which security goals each scenario evaluates.

**Adversary classes.** - **A1 — Malicious verifier.** A verifier that follows the protocol to receive presentations but seeks to (i) extract more PII than disclosed, (ii) replay or re-present credentials it has received, or (iii) collude with other verifiers to link a holder's interactions across services. - **A2 — Semi-honest issuer.** An issuer that issues credentials honestly but seeks to track holder usage of those credentials post-issuance, to issue equivalent credentials to unintended subjects, or to register credential types outside its domain of authority. - **A3 — Semi-honest government relay.** The government acts as the relay (§5.2 Phase 2) and follows the protocol, but seeks to maximize metadata observation while remaining content-blind to credential payloads. - **A4 — Network observer.** An attacker observing on-chain queries (chain-read traffic, RPC endpoints) who seeks to identify which verifier queries which DID. We assume the attacker does not control consortium validators.

Verifiers that operate their own read-only Guoxin Chain full node (see §4.2) sit outside this attack surface entirely, since their queries never leave their infrastructure. - **A5 — Consortium-majority adversary.** A coalition of more than the consensus threshold of consortium validators that seeks to write malicious entries to the DID Registry, Template Registry, or Revocation Registry. We treat this as a worst-case bound rather than a routine threat.

**Security goals.** We aim for: - **G1 — Issuer authentication.** Verifiers can cryptographically confirm that a credential was issued by a named, on-chain-resolvable issuer (against A1, A2). - **G2 — Holder authentication.** Verifiers can confirm that the presenter currently controls the DID-bound private key (against A1). - **G3 — Selective disclosure of contents.** Holders disclose only the claim fields they choose; verifiers learn nothing about omitted fields beyond what is structurally implied (against A1). - **G4 — Content blindness at the relay.** The relay cannot read credential payloads (against A3, at the payload level). - **G5 — Verifier privacy.** No central authority observes which verifier queries which DID at the application layer (against A3, A4 modulo network-level metadata). - **G6 — Accountability under legal process.** The state retains a defined, gated capability to deanonymize a DID through the linkage table.

The architecture targets G1–G6. It does *not* target presentation-unlinkability (defeated by single-DID reuse, see §6.4), credential-graph privacy against the relay (defeated by per-issuance metadata logging, see §5.5), or network-layer verifier anonymity (verifier IPs are visible to whoever hosts the chain RPC). These omissions are named explicitly so that the architecture's claims are evaluated on what it actually delivers.

**Security claims: mechanism-to-goal mapping.** Table 1 summarises how each security goal is realized by a specific architectural mechanism, which adversary classes the mechanism addresses, and what assumptions or limitations bound the claim.

| Security Goal | Architectural Mechanism | Adversary Classes | Assumptions & Limitations |
|---|---|---|---|
| **G1** — Issuer authentication | Issuer signature on VC (Phase 2 step 2); verifier resolves issuer DID document on-chain and validates signature (Phase 3 step 7) | A1 (malicious verifier), A2 (semi-honest issuer seeking to issue equivalent credentials to unintended subjects) | Issuer private key remains uncompromised; DID document integrity is guaranteed by consortium consensus; verifier correctly resolves the issuer DID |
| **G2** — Holder | Holder signature | A1 (malicious | Holder private key |

| Security Goal | Architectural Mechanism | Adversary Classes | Assumptions & Limitations |
| --- | --- | --- | --- |
| authentication | on presentation (Phase 3 step 6); verifier validates against DID document public key (Phase 3 step 7); verifier-provided nonce + timestamp in presentation prevent replay | verifier attempting replay) | remains uncompromised; verifier includes fresh nonce in request; no key-loss recovery mechanism specified (§7.4) |
| **G3** — Selective disclosure of contents | SD-JWT with hash-committed disclosures: issuer includes `_sd` hash commitments for selectively-disclosable fields in the signed VC; holder reveals only chosen fields plus salts at presentation (Phase 3 step 6); verifier validates disclosed values against issuer-signed commitments (Phase 3 step 7); per-field `selective_disclosable` flag in template schema determines which fields are commitment-protected (§5.4) | A1 (malicious verifier seeking excess PII) | Underlying hash function is preimage-resistant; SD-JWT does not by itself provide presentation unlinkability — single-DID reuse across presentations still enables verifier collusion to link interactions (§6.4) |
| **G4** — Content blindness at relay | Issuer encrypts VC payload to holder's public key before | A3 (semi-honest relay) | Issuer performs encryption to the correct holder |

| Security Goal | Architectural Mechanism | Adversary Classes | Assumptions & Limitations |
|---|---|---|---|
| | transmission (Phase 2 step 2); relay sees ciphertext only, holds no decryption key (Phase 2 step 3); separate key-agreement key mitigates cross-protocol key-use risk (§5.3) | | public key; underlying encryption scheme (HPKE/SM2) is IND-CCA2 secure; issuer and holder key-agreement keys are correctly published in DID documents |
| **G5** — Verifier privacy | Public-readable chain with anonymous queries (§4.1–4.2); no authentication, API key, or permission required for DID resolution or signature validation; verifier-operated read-only full node option eliminates third-party RPC observation (§4.2) | A3 (semi-honest relay), A4 (network observer) | For verifiers using shared consortium RPC endpoints, the RPC operator sees query metadata (verifier IP, queried DID); network-layer traffic analysis (by ISP or state-level observer) is outside the architecture's threat model; chain-level read patterns remain publicly visible |
| **G6** — Accountability under legal process | Off-chain linkage table (`CTID identity ↔ did:sic:`) held exclusively by the state (§4.3 step 4); access gated by judicial warrant or equivalent legal process; storage protected by access controls, | — (affirmative capability, not a defense against a specific adversary) | Linkage-table compromise (§6.4) is the single highest-value attack surface; legal-process integrity is a governance rather than cryptographic guarantee; deanonymization capability is total |

| Security Goal | Architectural Mechanism | Adversary Classes | Assumptions & Limitations |
|---|---|---|---|
| | encryption, and audit logging (§4.5) | | for any targeted DID |

*Table 1: Security claims mapped to architectural mechanisms, adversary classes, and assumptions.*

## 3.1 Verifier-Side PII Burden and Data Locality

**Scenario.** A user wishes to access a third-party corporate-benefits platform — for example, a discount portal that partners with employers to offer perks to their staff. The platform must verify two facts about the user: (i) their legal identity (to prevent fraudulent signups) and (ii) that they are currently employed at one of the participating companies.

**What the centralized model requires.** Two friction layers compound. *First*, to verify legal identity via CTID, the platform must already hold the user's name, identification number, and facial data (per §2.1, CTID expects these as query inputs and returns only yes/no), then submit them to CTID. The platform becomes a custodian of full PII for every prospective member, regardless of which company they actually work for. *Second*, employment status is not in CTID — CTID anchors *legal identity*, not employment relationships. The platform therefore must additionally obtain employment confirmation from each participating employer's HR system. With no public HR API, this means either bespoke per-employer integrations (which do not scale across thousands of partner companies) or informal email/paper attestations (which create new fraud and leakage surfaces).

**Threat model.** Two actors carry risk. The platform, having collected full PII from every user just to bootstrap CTID verification, becomes a high-value breach target. Each employer's HR system, queried piecemeal by every third-party benefits provider it partners with, becomes an attack surface and a structural leak point — and there is no clean way for an HR system to confirm employment to an external verifier without disclosing more than necessary. In the model of §3.0, the platform embodies **A1** (malicious verifier), structurally compelled to collect PII far beyond what the transaction requires; the harm it creates is precisely what **G3** (selective disclosure of contents) is designed to prevent.

**RealDID resolution.** The employer (a non-government entity that has registered as an issuer on the open Template Registry and published an `EmploymentCredential` template — see §5.1) issues a VC to each employee at onboarding, encrypted to the employee's public key and counter-signed by the government relay. To access the benefits platform, the employee presents only the fields the platform actually needs — for example, `{employer: "ABC Corp", status: "active"}` — omitting their department, title, hire date, and identity. The platform validates the triple-signature chain on the open chain, confirms the issuer DID resolves to the named employer, and confirms the holder's signature against the holder DID, all without collecting any PII and without any direct query to the employer's HR system. The platform never becomes a PII custodian; the employer's HR data never leaves the employer's perimeter.

This scenario also illustrates an aspect of the architecture made explicit in §5.1: any qualifying entity — government bureaus, universities, hospitals, employers, professional bodies, financial institutions — can register as an issuer and define its own credential template, subject to the government's approval that the issuer is authoritative for the credential type. The set of issuable credentials is not limited to government-held data.

## 3.2 Cross-Border Credential Verification

**Scenario.** A Chinese university graduate applies to a foreign university for a master's program. The foreign university requires verification that the applicant's bachelor's degree is authentic and was issued to a real person. The university has no access to CTID and no standing to query China's national identity infrastructure.

**What CTID requires (and why it fails across borders).** Two prerequisites fail simultaneously. *First*, the foreign university would need to already hold the graduate's PII (per §2.1, CTID takes name + identification number + facial data as inputs, not outputs) — possible in principle, but forcing the foreign verifier to become a custodian of Chinese citizens' identity data. *Second*, the foreign university would need network access to the CTID endpoint, which is not exposed across borders. The graduate therefore must rely on notarized paper documents, third-party credential evaluation services, or bilateral institutional agreements, all of which introduce delay, cost, and trust dependencies.

**Threat model.** The verifier (foreign university) cannot distinguish between a genuine graduate with authentic credentials and an applicant presenting forged documents. The credential evaluation service becomes a centralized trust bottleneck. The graduate's personal data crosses borders through multiple intermediaries, each a potential leak vector. In the model of §3.0, the foreign university's inability to distinguish genuine credentials from forgeries is a failure of both **G1** (issuer authentication) and **G2** (holder authentication). The PII exposure through intermediaries violates **G3** (selective disclosure), and the structural exclusion of foreign verifiers from the CTID API is an instance of the access inequality that **G5** (verifier privacy) addresses at the architectural level.

**RealDID resolution.** The graduate's university (an authorized issuer) has registered a DiplomaCredential template on the RealDID chain. The graduate holds a VC containing their degree, major, and graduation date, encrypted to their private key, with issuer and government counter-signatures. The foreign university queries the Chinese permissioned chain — publicly readable by anyone anywhere — resolves the issuer's DID, validates the three signatures, and confirms the credential is authentic. The foreign university never accesses CTID, never learns the graduate's national ID number, and never touches China's identity infrastructure. The graduate's PII never crosses the border.

## 3.3 Anonymous Accountability

**Scenario.** A user wishes to participate in an online medical Q&A community as an answering practitioner — answering health questions posed by anonymous askers. The community requires that every answering account be backed by a real, currently-licensed medical professional (to maintain answer quality and to meet platform-level liability commitments), but promises that practitioners can remain anonymous to askers and to the wider public.

**What the centralized model requires.** Two-step verification, each step over-collecting. *First*, the platform must verify the practitioner's legal identity through CTID — which means collecting name, identification number, and facial data (per §2.1) and submitting them to CTID for a yes/no confirmation. The platform now holds full PII for every practitioner. *Second*, the platform must verify the practitioner's current medical license. CTID does not hold license data, so the platform must additionally query the relevant licensing authority's database — assuming a queryable API exists for that jurisdiction at all. The end state: the platform possesses each practitioner's full legal identity bound together with their professional license record. The platform's promise of practitioner anonymity is contradicted by the platform's structural possession of identity.

**Threat model.** The platform is a semi-honest actor — it intends to keep practitioner identities private from askers and from the public, but holds identity-plus-license data that can be compelled, breached, or misused by insiders. The threat is not the platform's intent but its structural possession of the data. In the model of §3.0, the platform's structural possession of identity-plus-license data is the threat that **G3** (selective disclosure of contents) eliminates: the verifier receives only the claims it needs — `{license_valid: true, specialty: "general practice"}` — and never touches the practitioner's identity. The government's retained linkage capability implements **G6** (accountability under legal process) without requiring continuous identity exposure to every verifier.

**RealDID resolution.** The relevant licensing authority (a non-government issuer in the sense of §5.1 — a professional regulator that has registered as an issuer on the open Template Registry) publishes a `MedicalLicenseCredential` template. A licensed practitioner holds a VC stating, for example, `{license_valid: true, specialty: "general practice", jurisdiction: "Beijing"}`, encrypted to their private key and counter-signed by the government relay. To participate as an answerer on the platform, the practitioner presents only `{license_valid: true, specialty: "general practice"}` — sufficient for the platform to confirm the answerer is a real, currently-licensed practitioner in the appropriate specialty. The platform validates the triple-signature chain and accepts the presentation; it never collects the practitioner's name, identification number, photograph, or license number. The practitioner is anonymous to the platform; the platform's quality bar is preserved. The government retains the RealDID ↔ CTID linkage, accessible only through defined legal process — providing accountability without continuous identity exposure.

This scenario demonstrates that the architecture supports *verified-but-anonymous* participation not only at the level of identity uniqueness ("a real, unique person") but also at the level of professional or institutional standing ("a real, currently-licensed practitioner"). The credential layer carries the claim the verifier actually needs; the identity layer remains undisclosed.

## 3.4 Centralized API Cost and Verifier Access Inequality

**Problem.** CTID's single permission-based API gateway creates two compounding failures: (1) **infrastructure cost** — the government bears the full operational expense of handling national-scale verification traffic, a burden that grows with digital service adoption; (2) **verifier access inequality** — only a limited set of authorized entities may verify identities. Small businesses, educational institutions, non-profits, and foreign organizations are structurally excluded. The digital identity ecosystem is restricted to a small, privileged class of verifiers. In the model of §3.0, the centralized API creates the observation surface that **A3** (semi-honest government relay) exploits — every verification request passes through a single point — and eliminates **G5** (verifier privacy) as a dimension of the identity ecosystem.

**RealDID resolution.** Verification moves from the government's API gateway to the open permissioned chain. Any party with network access can query DIDs, resolve DID documents, and validate cryptographic signatures — no authorization required. The government eliminates the operational cost of the verification API. The verifier base expands from a handful of authorized entities to the entire internet-connected world. The chain's consensus mechanism distributes the cost of maintaining the verification infrastructure across the consortium.

## 3.5 Verifier Privacy

**Problem.** In the centralized model, every verification request passes through the government's API gateway. The government observes: who is verifying whom, when, how often, and in what sequence. A bank verifying a customer, a hospital verifying a patient, a school verifying a parent — the government records all of it. In an identity ecosystem, *the identities of verifiers* and *the relationships between verifiers and subjects* are themselves sensitive privacy data. A company's customer list, a clinic's patient roster, a law firm's client base — all are exposed to the government through the verification API. This is not a bug; it is an architectural inevitability of centralized verification. In the model of §3.0, the government's complete visibility into verification transactions is **A3** (semi-honest government relay) operating exactly as the centralized architecture permits: every API call is an observation event. The harm is the total loss of **G5** (verifier privacy).

**RealDID resolution.** At the protocol level, on-chain verification creates no central observation point: a verifier resolves a DID document and validates a signature without authenticating to any authority, calling any API, or obtaining any permission. There is no structural equivalent of the CTID gateway that logs every verification request. The architecture therefore provides a *spectrum* of verifier privacy under the verifier's own control. Verifiers with sufficient operational capacity can operate their own read-only Guoxin Chain full node and perform DID resolution and signature validation entirely on their own infrastructure, with no query ever leaving the verifier's network. In that mode, even the consortium nodes — the operators of the chain itself — cannot observe which DIDs the verifier is checking, when, or in what sequence. This realizes **G5** (verifier privacy) against both **A3** (semi-honest government relay) and **A4** (network observer): at the application layer, verification leaves no trace at any central authority. Verifiers without that capacity may query a shared consortium RPC endpoint — a choice that trades some privacy (the RPC operator sees the verifier's IP and the queried DID) for operational simplicity, but which the architecture does not require. The key architectural property is that the *protocol itself* creates no observation point: the decision is the verifier's, not the system's. Verifier privacy emerges as a structural property of the architecture — not a policy promise — but its strength depends on the verifier's operational choice. This protects an under-appreciated dimension of the identity ecosystem: the right of verifiers to verify without being tracked.

## 4. RealDID Architecture

The scenarios in §3.1–§3.5 illustrate five structural problems that centralized identity verification creates: verifier-side PII burden, cross-border exclusion, the tension between verification and anonymity, infrastructure cost concentration, and loss of verifier privacy. The RealDID architecture addresses each through a combination of an open permissioned chain (§4.1–4.2), a DID issuance protocol that anchors legal identity without exposing it on-chain (§4.3–4.5), a content-blind government relay, and a VC ecosystem with cryptographic selective disclosure (§5). This section presents the identity anchor layer. Figure 1 shows the three-layer architecture stack at a glance.

## Three-Layer Architecture

*CTID → RealDID → VCs — a legally-anchored decentralized identity stack*

### Architecture Stack

**LAYER 3 — Verifiable Credentials**

Rich, use-case driven credentials: diplomas, licenses, age proofs, certifications
Triple-signature chain: Issuer → Government → Holder
Selective disclosure, issuer-defined templates, encrypted content

**Government sees:** VC existence (not contents)
**Verifier sees:** disclosed claims only
**Data richness:** HIGH

**LAYER 2 — RealDID (did:realdid:)**

Decentralized identity anchor on open permissioned chain
Government issues DID after CTID verification
User-generated key pair — government has no private key access

**Government sees:** RealDID ↔ CTID linkage
**Verifier sees:** DID + public key on chain
**Data richness:** MINIMAL

**LAYER 1 — CTID (Centralized Trusted Identity)**

National online identity authentication system
Facial recognition, ID document verification, full PII
Centralized — operated by Ministry of Public Security

**Problems:** Single permissioned API gateway — enormous infrastructure cost, limited verifier access, government observes every verification

**Government sees:** Full PII + who verifies whom
**Verifier sees:** Nothing (not exposed)
**Data richness:** FULL

← Centralized    Semi-decentralized    User-sovereign →

### Key Architectural Guarantees

| Guarantee | How It's Achieved |
|---|---|
| User key sovereignty | Key pairs generated by users; government never holds private keys |
| Linkage secrecy | RealDID ↔ CTID mapping held exclusively by government; on-chain queries are anonymous |
| VC content blindness | Government encrypts VCs to holder's public key; authenticates participants without seeing contents |
| Selective disclosure | Holders present only required claims; template marks which fields are hideable |
| Verifier privacy | On-chain verification is anonymous; government cannot observe who verifies which DID |
| Infrastructure cost shift | Verification moves from government-operated API gateway to distributed on-chain queries |
| Verifier access equality | Anyone with chain access can verify; no permission-based gatekeeping |
| Anti-impersonation | Government can deanonymize (linkage table) but cannot sign as holder (no private key) |

### Centralized vs. Decentralized Verification

| Dimension | CTID Centralized Verification | RealDID Decentralized Verification |
|---|---|---|
| Verification gateway | Single government-operated API | Open permissioned chain — anyone can query |
| Infrastructure cost | Government bears full cost of national-scale traffic | Distributed across chain validator nodes |
| Verifier access | Permission-based — very limited entities | Permissionless — anyone with chain access |
| Verifier privacy | Government sees who verifies whom | On-chain queries are anonymous — government cannot observe |
| Data disclosed to verifier | Full PII (name, ID number, photo) | Only disclosed claims selected by holder |

*Figure 1: Three-layer architecture stack. CTID provides the centralized legal-identity root; RealDID anchors decentralized identifiers on the open permissioned chain; the VC layer carries application-level credentials with selective disclosure.*

## 4.1 Why Open Permissioned Chain

RealDID's chain architecture is shaped by three structural constraints that apply in China's legal and regulatory environment.

**Cryptocurrency prohibition.** Public permissionless blockchains (Ethereum, Solana, etc.) require cryptocurrency for gas fees to execute transactions and store state. Cryptocurrency transactions are legally restricted in China, making public permissionless chains unavailable as national infrastructure. A private single-entity chain avoids this issue but defeats the purpose of decentralized verification: if one party controls the ledger, verifiers must trust that party to not tamper with DIDs, templates, or revocation entries.

**API cost elimination.** CTID's centralized verification gateway concentrates infrastructure cost on the government. Moving verification to a distributed chain shifts operational costs to the consortium of validator nodes, eliminating the single-bottleneck API as a cost center. Verifiers with sufficient infrastructure can take this further by operating their own read-only Guoxin Chain full node and bearing their own verification-side compute and storage costs — neither the government nor the consortium pays for those queries. Small verifiers without that capacity simply query a public consortium RPC endpoint at no cost.

**Verifier access equality.** A permissioned API restricts verification to authorized entities. A publicly readable chain enables any party to verify DIDs and signatures without permission, expanding the verifier base from a privileged few to anyone with network access.

The **open permissioned chain** satisfies all three constraints: permissioned validators (consortium members) avoid cryptocurrency dependence; public readability enables permissionless verification; and multi-stakeholder governance prevents unilateral control — the architectural defense against **A5** (consortium-majority adversary). No single entity, including the state, can unilaterally alter the registries. This is the closest architecture to a public permissionless chain that is legally permissible in China.

## 4.2 Chain Topology

The RealDID chain is a consortium blockchain with the following properties:

- **Validator set.** Permissioned nodes operated by consortium members — government agencies (such as national identity and cybersecurity authorities) and private sector entities (technology platforms, telecommunications carriers, financial institutions).

- **Public read access.** Any party with network access can query the chain: resolve DID documents, read the template registry, check the revocation registry, and validate signatures. No authorization, API key, or permission is required. Verifiers may either query a consortium-hosted RPC endpoint or operate their own read-only full node (sync against the public chain state); the chain's read interface is open at the protocol level, not gated by the consortium.
- **Smart contract governance.** Three smart contracts govern the identity infrastructure: the DID Registry (issuance, update, deactivation of DIDs), the Template Registry (registration, approval, deprecation of VC templates), and the Revocation Registry (credential revocation, DID deactivation). All three are governed by consortium consensus — no single entity, including the government, can unilaterally modify them. This is the enforcement mechanism for the defense against **A5** (consortium-majority adversary): malicious registry entries require collusion above the consensus threshold (§6.4).
- **Consensus.** The PBFT (Practical Byzantine Fault Tolerant) consensus protocol, as implemented in the FISCO BCOS open-source blockchain framework [33], suitable for a permissioned validator set with known identities.

### 4.3 DID Issuance Protocol

The protocol for issuing a RealDID proceeds in five steps:

**Step 1: CTID Verification.** The user invokes CTID's per-session face+ID match against their pre-existing CTID record (which was populated automatically when they obtained their national ID card — see §2.1). CTID confirms that the submitted live face image and identification number match the on-file biometric and documentary data, returning a yes/no authentication result. No separate CTID registration is required at this point — only a per-session verification against an already-populated record. This step establishes the user as a verified legal person at the moment of DID issuance.

**Step 2: Key Generation.** The user's device generates a cryptographic key pair locally. The public key is transmitted to the government; the private key remains exclusively on the user's device and is never transmitted to any party.

**Step 3: DID Registration.** The government submits a transaction to the DID Registry smart contract, registering `did:sic:<address>` with the user's public key. The consortium validates and commits the transaction. The DID document is now publicly resolvable on-chain.

**Step 4: Linkage Recording.** Internally, the government records the mapping `CTID identity ↔ did:sic:<address>` in a secure, off-chain linkage table. This mapping is never published on-chain. It is accessible only to the government, and only for purposes defined by law (e.g., criminal investigation with judicial warrant). This step implements **G6** (accountability under legal process): the state retains a defined, gated deanonymization capability while keeping the linkage off-chain, invisible to **A4** (network observer).

**Step 5: Key Custody.** The user retains exclusive control of the private key. The government has no copy, no escrow, and no mechanism to sign as the user. The DID is now operational: the user can sign VC applications, receive encrypted VCs, and present credentials to verifiers.

## 4.4 Key Custody and the Asymmetric Trust Model

The key custody design establishes an asymmetric, state-bounded relationship between the state and the user, summarized in Table 2. The rows of the table correspond to the security goals of §3.0: the government's inability to prove control of a DID, view VC contents, or impersonate a subject realizes **G4** (content blindness at the relay) against **A3**; its retained deanonymization and revocation capabilities implement **G6** (accountability under legal process).

| Capability | Government | User |
|---|---|---|
| Prove control of DID | No (no private key) | Yes (holds private key) |
| Deanonymize a DID | Yes (linkage table) | No |
| Fabricate legal identity | No (requires CTID verification) | No (requires CTID verification) |
| View VC contents | No (encrypted to holder's public key) | Yes (decrypts with private key) |
| Impersonate a DID subject | No (no private key) | N/A (is the subject) |
| Revoke a DID | Yes (DID Registry) | No |

*Table 2: Asymmetric, state-bounded capabilities. The state retains broader structural powers (linkage, revocation, metadata observation) than the user, but is cryptographically excluded from impersonation and from reading credential contents. Neither party can act unilaterally across all dimensions, but the constraints are not symmetric.*

**Asymmetric-constraint property.** The state cannot impersonate, cannot read credential contents, and cannot unilaterally alter the registries. The user cannot fabricate a legal identity and cannot prevent lawful deanonymization, revocation, or metadata observation by the relay. These constraints are real but not symmetric: the user is bounded by *cryptography*, while the state is bounded by *cryptography plus legal-process commitments around capabilities it both holds and operates*.

The system is not "trustless" in the blockchain sense, nor a clean two-way balance of power — it is *trust-distributed with a structurally privileged state root*. Throughout this paper we use the term **decentralized verification with centralized legal anchoring** to describe this arrangement. The policy implications of the asymmetry are examined in §6.2.

### 4.5 Linkage Secrecy

The RealDID ↔ CTID linkage table is the most sensitive component of the architecture. It is held exclusively by the government, stored off-chain, and protected by access controls, encryption, and audit logging. On-chain DIDs reveal no information about the underlying legal identity: a DID document contains only a public key and service endpoints. An observer monitoring the chain cannot associate a DID with a real-world person.

Deanonymization requires a judicial warrant or equivalent legal process. The linkage table is not a general-purpose lookup service; it is accessible only through defined legal procedures. This design provides accountability to law without enabling casual surveillance. It is the architectural implementation of **G6**: deanonymization is possible but gated, and an **A4** (network observer) monitoring the chain learns nothing — DIDs are opaque strings with no on-chain link to real-world identity.

**Deployment status.** The architectural description in this section reflects the RealDID system as designed and deployed. Operational metrics will be included in a future revision; see the Working Paper notice above.

## 5. VC Ecosystem on RealDID

With RealDID providing the decentralized identity anchor, verifiable credentials form the rich, use-case-driven layer on top. This section describes the VC template registry, the full credential lifecycle, the triple-signature chain, credential schemas, and the content-blindness properties that protect holder privacy.

### 5.1 VC Template Registry

Before any credential can be issued, issuers must register VC templates on-chain. A template defines a credential type, its claim schema, the business scenario it serves, and usage constraints. This registry serves as a catalog: holders browse available templates to discover which credentials they may apply for, and verifiers consult templates to understand the structure and semantics of credentials they receive.

**Open issuer model.** The issuer set is not restricted to government agencies. Any qualifying entity — government bureaus, universities, hospitals, employers, professional bodies, financial institutions, and other social and industrial organizations — can register as an issuer by submitting templates for the credential types it is authoritative for. A hospital can issue `VaccinationCredential` or `PatientCredential`, an employer can issue `EmploymentCredential`, a trade association can issue `ProfessionalMembershipCredential`, a bank can issue `AccountStatusCredential`, and so on. The government's role at template registration is not to be the issuer itself but to gate participation: it verifies, through the KYC and manual approval process described below, that each submitting issuer is authoritative for the credential type it proposes.

The template registration process (Phase 0) proceeds as follows:

1. An issuer creates a template specifying: credential type (e.g., `DiplomaCredential`), claim schema (field names, types, whether each field is required, and which fields support selective disclosure), business scenario description, and constraints (e.g., maximum validity period).
2. The template is submitted to the government for approval. Issuers undergo a KYC process to establish their legal identity and organizational authority; each VC template is then reviewed and approved manually by human operators, who verify that the submitting issuer is authoritative for the credential type it proposes — a university for `DiplomaCredential`, a medical board for `ProfessionalMedicalLicense`.
3. Upon approval, the template is published to the on-chain Template Registry with the government's approval signature and a timestamp. Holders can now discover the template and apply for VCs against it.

## 5.2 VC Lifecycle

The full VC lifecycle spans eight steps across four phases, with a preceding one-time **Phase 0** for issuer-driven template registration. Figure 2 illustrates the flow; the Government's role is confined to Phase 2 (authenticate participants, log metadata, counter-sign — without reading credential contents) and to revocation in Phase 4. Detailed exposition of each step follows below.

## VC Lifecycle on RealDID

*One-time template registration (Phase 0) + per-credential 8-step flow (Phases 1–4)*

**PHASE 0 — TEMPLATE REGISTRATION (ISSUER-DRIVEN, ONE-TIME PER CREDENTIAL TYPE)**

ISSUER
**⓪-a Template Register**
Issuer creates VC template
Defines credential type, claim schema, business scenario, and usage constraints.

→

GOVERNMENT
**⓪-b Gov Approve**
Government approves template
Verifies issuer is authorised to issue this credential type. Template published on-chain.

→

ON-CHAIN
**⓪-c Discoverable**
Templates available on-chain
Holders browse the Template Registry before applying for VCs.

**PHASE 1 — APPLICATION & ISSUANCE**

HOLDER → ISSUER
**① Apply**
Holder applies for VC
Selects template; signs application with private key; proves DID control.

→

ISSUER → GOV. RELAY
**② Issue**
Issuer creates VC
Populates template with claims; encrypts to holder's public key; signs with issuer DID.

**PHASE 2 — GOVERNMENT GATE** — CONTENT-BLIND RELAY

GOVERNMENT
**③ Verify**
Gov authenticates participants
Checks (a) issuer authorisation; (b) holder DID active. **Cannot read payload.** Logs metadata: `{issuer DID, holder DID, template ID, VC hash, timestamp}`.

→

GOV. → HOLDER
**④ Deliver**
Gov counter-signs & relays
Adds counter-signature attesting both parties are authentic; relays the encrypted VC to the holder.

**PHASE 3 — STORAGE, PRESENTATION, VERIFICATION**

HOLDER
**⑤ Store**
Holder stores VC
Decrypts with private key; stores in digital wallet.

→

HOLDER → VERIFIER
**⑥ Present**
Holder presents with selective disclosure
Constructs presentation with chosen claims; signs with private key.

→

VERIFIER
**⑦ Verify**
Verifier performs four checks
(a) holder signature, (b) issuer signature, (c) government counter-signature, (d) Revocation Registry lookup.

**PHASE 4 — REVOCATION**

ISSUER OR GOV.
**⑧ Revoke**
Publish to Revocation Registry
Issuer revokes individual VCs (e.g., licence expiry); Government revokes whole RealDIDs (e.g., fraud, legal process). Cascades to dependent VCs. Verifiers check this registry as part of step ⑦.

*Figure 2: VC lifecycle on RealDID. Phase 0 (purple) registers credential templates on-chain — a one-time issuer-driven step. Phases 1–4 (blue / pink-dashed / green / amber) describe the per-credential flow from application through revocation. The dashed border on Phase 2 highlights the Government's content-blind relay role: it authenticates issuer and holder and logs metadata, but cannot read the encrypted payload.*

**Phase 1 – Application and Issuance.**

*Step 1: Apply*. The holder selects a template from the registry and submits a VC application to the issuer. The application is signed with the holder's private key, proving control of the DID and authenticating the request.

*Step 2: Issue*. The issuer validates the application against the template, populates the template with the appropriate claims, encrypts the resulting VC to the holder's public key, signs it with the issuer's DID private key, and submits it to the government relay. Encryption at the issuer — before the credential reaches the relay — is the mechanism that realizes **G4** (content blindness at the relay) against **A3**: the relay will see only ciphertext.

**Phase 2 – Government Gate.**

*Step 3: Verify*. Before the VC reaches the holder, it passes through the government relay. The government verifies two facts: (a) the issuer is authorized to issue this credential type, and (b) the holder DID corresponds to a valid, active RealDID. Critically, the government cannot view the VC contents — the payload was encrypted *by the issuer* in Step 2 above to the holder's public key, and the government holds no decryption key. Encryption is performed at the issuer, not at the relay; the relay sees only ciphertext. The government logs metadata – `{issuer DID, holder DID, template ID, VC hash, timestamp}` – for audit purposes.

*Step 4: Deliver*. The government adds a counter-signature attesting that both parties are authentic, and relays the encrypted VC to the holder.

**Phase 3 – Storage, Presentation, and Verification.**

*Step 5: Store*. The holder receives the encrypted VC, decrypts it with their private key, and stores it in their digital wallet. The VC is now ready for use.

*Step 6: Present*. When a verifier requests proof of certain claims, the holder constructs a presentation containing the issuer-signed SD-JWT together with disclosure objects — claim name, value, and salt — for only the claims the verifier requires. The verifier can hash each disclosed value-plus-salt and validate it against the `_sd` commitments in the issuer-signed credential, cryptographically confirming that the disclosed claims originated from the issuer. Claims marked as selective-disclosable in the template are commitment-protected; required claims need not carry `_sd` entries. The holder signs the presentation with their private key.

*Step 7: Verify*. The verifier receives the presentation and performs five checks: (a) resolves the holder's DID document on-chain and validates the holder's signature against the public key; (b) resolves the issuer's DID document and validates the issuer's signature on the original SD-JWT; (c) validates the government's counter-signature; (d) for each disclosed claim, hashes the value-plus-salt and validates the result against the `_sd` commitments in the issuer-signed SD-JWT — confirming the claim originated from the issuer without revealing undisclosed fields; (e) checks the on-chain Revocation Registry to confirm neither the VC nor the holder's DID has been revoked. Checks (a) and (b) realize **G2** (holder authentication) and **G1** (issuer authentication) respectively — against **A1** (malicious verifier attempting replay or accepting forged credentials) and **A2** (semi-honest issuer issuing to unintended subjects). Check (d) realizes **G3** (selective disclosure of contents) cryptographically: the verifier confirms that disclosed claims are issuer-authenticated without learning the undisclosed claims. If all five checks pass, the verifier accepts the presentation.

**Phase 4 – Revocation.**

*Step 8: Revoke*. A VC may be revoked by its issuer (e.g., a license expires or is withdrawn) or the government may revoke all VCs associated with a DID (e.g., if the underlying RealDID is deactivated due to identity fraud or legal process). Revocation entries are published to the on-chain Revocation Registry. Verifiers are expected to check this registry as part of Step 7.

## 5.3 Triple-Signature Chain

Every VC presentation carries three cryptographic signatures, each serving a distinct trust function.

**Issuer Signature.** The issuer signs the VC at creation. This signature proves to the verifier that the credential was issued by an authoritative source – for example, that a diploma VC was genuinely issued by the university, not fabricated by the holder. The verifier validates this signature against the issuer's public key in the issuer's on-chain DID document. This is the mechanism that realizes **G1** (issuer authentication) against **A1** (malicious verifier) and **A2** (semi-honest issuer).

**Government Counter-Signature.** The government signs the VC during the relay (Phase 2, Step 4). This signature proves two facts simultaneously: the issuer is authorized to issue this credential type, and the holder is a verified RealDID subject. The government does not attest to the truth of the claims – only to the authenticity of the participants. The verifier validates this signature against the government's public key, which is published on-chain. This signature bridges the gap between **G1**/**G2** and the state's legal-anchoring function: it cryptographically binds the government's attestation of participant authenticity to the credential without requiring the government to see credential contents (**G4**).

**Holder Signature.** The holder signs each presentation at the moment of disclosure. This signature proves that the presenter currently controls the private key associated with the DID – that the VC is being presented by its legitimate owner, not replayed by a third party. With a verifier-provided nonce and timestamp included in the presentation, this is the mechanism that realizes **G2** (holder authentication) against **A1** (malicious verifier attempting replay).

All three signatures must validate for a presentation to be accepted. A missing or invalid issuer signature means the credential is not authoritative. A missing or invalid government signature means the participants' authenticity is unverified. A missing or invalid holder signature means the presenter is not the credential's owner or the presentation has been tampered with.

**Cryptographic primitives.** The architecture is parameterized over the underlying primitives rather than tied to a specific suite, and the choice of suite is constrained by the deploying jurisdiction's cryptographic regulations. In China specifically, financial and identity-grade systems are typically required to use the SM-series national standards (SM2 for elliptic-curve signatures and key agreement, SM3 for hashing, SM4 for symmetric encryption); a RealDID deployment in China is expected to use these. In illustrative terms — and abstracting over the specific suite — the architecture requires: (i) an EUF-CMA-secure signature scheme such as SM2 or Ed25519 [13] applied over a canonicalized representation of each signed object (issuer signature over the VC instance; government counter-signature over `H(VC instance) || issuer DID || holder DID || template ID || timestamp`; holder signature over the presentation including a verifier-provided nonce and timestamp to prevent replay), and (ii) a hybrid public-key encryption scheme — in the SM-suite case based on SM2/SM3/SM4, or in other jurisdictions a construction such as HPKE [14] over X25519 with AES-256-GCM — for content encryption from issuer to holder. Following W3C DID core privacy guidance [1], DID documents should publish *separate* verification methods for signing and for key agreement so that the holder's signing key (DID-bound, used for authentication) and the holder's key-agreement public key (used by issuers for content encryption) are not the same key in the same role.

The current deployment uses SD-JWT [12] for selective disclosure. The issuer includes hash commitments (`_sd`) for each selectively-disclosable claim in the signed VC; at presentation, the holder reveals only the chosen claims together with their per-claim salts. The verifier hashes each disclosed value-plus-salt and validates the result against the commitments in the issuer-signed credential. This provides cryptographic assurance that the undisclosed claims are issuer-authenticated — the holder cannot alter them — without revealing them to the verifier. Content blindness at the relay (**G4**) is unaffected: the relay counter-signs the encrypted SD-JWT as a single opaque object (§5.2 Phase 2) and does not interact with the selective-disclosure mechanism, which operates entirely at the holder–verifier presentation step. Substituting a scheme such as BBS+ [10] for the issuer signature, which would additionally provide presentation unlinkability at the cryptographic layer, remains compatible with the architecture and is a candidate for future iterations. We treat the specific primitive selection as a deployment parameter rather than a load-bearing architectural commitment.

## 5.4 Credential Schemas

The RealDID VC ecosystem uses three W3C-aligned schema structures: the template, the VC instance, and the presentation.

**Template Schema** (registered on-chain, Phase 0):

```
template_id: "did:sic:tsinghua#DiplomaCredential-v1"
issuer_did: "did:sic:tsinghua"
credential_type: "DiplomaCredential"
business_scenario: "Higher education degree verification"
claim_schema:
  degree:       { type: string,  required: true,  selective_disclosable:
 false }
  major:        { type: string,  required: true,  selective_disclosable:
 false }
  graduation:   { type: date,    required: true,  selective_disclosable:
 false }
  gpa:          { type: float,   required: false, selective_disclosable:
 true  }
  honors:       { type: string[], required: false, selective_disclosabl
e: true  }
constraints:
  max_validity_days: null
gov_approval_signature: <government signature>
registered_at: <timestamp>
```

**VC Instance** (issued to holder, encrypted with holder's public key):

```
vc_id: "urn:vc:0x7f3a9b..."
template_id: "did:sic:tsinghua#DiplomaCredential-v1"
holder_did: "did:sic:0x9b2c..."
claims:
  degree: "Bachelor of Engineering"
  major: "Computer Science"
  graduation: "2025-06-30"
  gpa: 3.82
issuer_signature: <issuer signature>
gov_counter_signature: <government signature>
issued_at: <timestamp>
expires_at: null
```

**Presentation** (holder to verifier, SD-JWT with selective disclosure):

```
vc_id: "urn:vc:0x7f3a9b..."
issuer_signed_sd_jwt: "<SD-JWT with _sd hash commitments, issuer signat
ure>"
disclosures:
  degree:       { value: "Bachelor of Engineering", salt: "0xa3f2..." }
  major:        { value: "Computer Science",        salt: "0xb4e1..." }
  graduation:   { value: "2025-06-30",              salt: "0xc5d0..." }
gov_counter_signature: <government signature>
holder_signature: <holder signature>
timestamp: <now>
nonce: <verifier-provided nonce>
```

In this example, the holder has chosen to disclose their degree, major, and graduation date — sufficient for an employer to verify qualification — while GPA and honors information remains cryptographically bound by the issuer's `_sd` hash commitments. The verifier can confirm that the undisclosed fields exist in the issuer-signed credential and are authentic, without learning their values.

## 5.5 Content Blindness and Role Visibility

The RealDID VC architecture enforces content blindness through encryption: VCs are encrypted to the holder's public key by the issuer, and the government relays the encrypted payload without decryption capability. This means the government authenticates participants in a transaction it cannot read. The resulting visibility matrix for all four roles is shown in Table 3.

| Data | Government | Issuer | Holder | Verifier |
|---|---|---|---|---|
| RealDID-CTID linkage | Full access | No access | No access | No access |
| VC existence (metadata) | Yes | For own VCs | All own VCs | Only when presented |
| VC contents | No (encrypted) | For own VCs | Full access | Disclosed claims only |
| Other issuers' VCs | No | No | Yes (if holder) | No |
| Verification queries | No (anonymous) | No | No | Own queries only |

*Table 3: Role visibility matrix. The government authenticates participants without viewing credential contents.*

The content-blindness property is a critical architectural guarantee: it means the government can vouch for the authenticity of participants in the VC ecosystem without becoming a surveillance point for the *content* of credential exchanges. The government knows that a VC was issued, by whom, to whom, and when — but not what it contains.

**Honest limit: the credential graph.** Content blindness is true for credential payloads but not for credential-graph metadata. Because the government relay logs `{issuer DID, holder DID, template ID, VC hash, timestamp}` for every issuance (§5.2 step 3), the state observes — across all users and all issuers — a complete bipartite graph of who holds which *types* of credentials, from whom, and when. This is the observation surface that **A3** (semi-honest government relay) exploits: the relay follows the protocol as specified, yet still accumulates, as a structural byproduct, a dataset whose re-identification potential the metadata-privacy literature has repeatedly demonstrated for analogous communication and transaction graphs [7, 8]. The architecture therefore protects *content* against the state but does not protect the *credential graph*. Mitigations such as template-ID obfuscation, blinded issuance, or oblivious-relay constructions are possible in principle [9] but are not part of the current RealDID design; we revisit this in §6.4 and §8.

## 5.6 Wallet Layer

The RealDID architecture does not mandate any specific wallet implementation. Instead, it specifies a wallet-provider policy that admits a plurality of wallets, with three properties.

**Multiple wallet providers.** Any application — mobile apps, mini-programs embedded in super-apps, browser extensions, or hardware devices — can apply to be a RealDID wallet provider. Approval requires demonstrating compliance with two architectural requirements: (i) the holder's private key never leaves the holder's device or is otherwise transmitted to the wallet provider's backend, and (ii) the wallet provider obtains explicit user consent before accessing the private key for any operation. Once approved, the wallet may discover templates from the on-chain Template Registry and hold encrypted and decrypted VCs on behalf of its users.

**Issuer-hosted wallets.** A consequence of the open-wallet policy is that an issuer can operate its own wallet for VCs it issues. In this configuration, the issuance and use flow is fully bilateral between issuer and holder: the issuer encrypts the VC to the holder's public key, the holder receives and decrypts it within the issuer-provided wallet, and no third-party wallet operator sits in the middle observing the interaction. This is particularly relevant for credential types where the issuer has a strong interest in being the sole observer of usage — for example, medical credentials issued by a hospital and consumed within the hospital's own patient portal.

**Privacy implication.** In single-wallet ecosystems — where one designated wallet app is the only legitimate holder of a given credential type — the wallet operator becomes a structural observation point: it sees which credentials the user holds, which verifiers they present to, when, and how often. RealDID's open-wallet policy distributes this observation surface across many independent wallet providers (none of whom hold private keys), and in the issuer-hosted case eliminates the third-party wallet operator entirely. We return to the contrast with single-wallet designs in §6.1.

# 6. Policy and Governance Analysis

This section analyzes the regulatory and governance implications of the RealDID architecture, examining alignment with Chinese data protection law, the dual-root trust model, accountability without mass surveillance, and the risks that must be honestly acknowledged.

## 6.1 Regulatory Alignment

**PIPL Data Minimization.** China's Personal Information Protection Law (PIPL) [6] establishes data minimization as a core principle: personal information collected must be limited to the minimum scope necessary to achieve the processing purpose, and excessive collection is prohibited. The centralized CTID model is in structural tension with this principle: every verifier is forced to collect and retain the subject's full PII in order to submit a verification query at all (§2.1), regardless of which single claim it actually needs to confirm. The RealDID architecture aligns with PIPL through two mechanisms. First, selective disclosure ensures that verifiers receive only the claims they require – an age threshold, a qualification, a professional license – not the holder's complete legal identity. Second, content blindness ensures that even the government, which mediates every VC issuance, sees only metadata (participant identity, timestamp, VC hash) and never the credential contents.

**Cross-Border Data Sovereignty.** PIPL imposes strict requirements on cross-border transfer of personal information, including security assessments, standard contracts, or certification. Under the centralized CTID model, cross-border verification is doubly blocked: a foreign verifier would need to (i) already possess the subject's PII to submit it as a CTID query input (forcing the foreign verifier to become a custodian of Chinese citizens' identity data) *and* (ii) have cross-border network access to the CTID endpoint, which is not exposed outside China. The first prerequisite displaces the PII-custody burden to a foreign jurisdiction; the second forecloses verification altogether. Under RealDID, foreign verifiers query the open permissioned chain directly. They resolve DIDs and validate signatures without accessing CTID, without receiving PII, and without routing requests through any Chinese government server. The verifier sees only the disclosed claims of a VC – which, in the DiplomaCredential example, is a degree title and graduation date, not a national ID number. Cross-border verification is achieved without cross-border PII transfer.

### *6.1.1 Comparison with eIDAS 2.0 and EUDI Wallet*

The European Union's eIDAS 2.0 regulation [3] and the accompanying European Digital Identity Architecture and Reference Framework (EUDI ARF) [15] provide the closest production-grade comparison point. Both RealDID and EUDI Wallet enable selective disclosure of credentials from a user-controlled wallet with cryptographic verifiability, and both use SD-JWT [12] at the credential-format layer (§5.3–5.4). The two architectures differ in two respects: trust-model anchoring and wallet design.

**Trust model.** eIDAS 2.0 distributes trust across a federation of member-state identity providers, while RealDID concentrates legal identity verification in a single state root (CTID) and distributes trust through the multi-stakeholder governance of the permissioned chain and smart contracts. Both models achieve user-controlled selective disclosure; they differ in how they anchor the legal identity that gives credentials their authority.

**Wallet architecture.** EUDI ARF prescribes a designated Personal Identification Data Wallet through which credentials are stored and presented within each member-state ecosystem. This concentrates a structural observation point at the wallet operator: the wallet application sees which credentials the user holds, which verifiers they interact with, and at what cadence — even when the underlying credentials are issued by independent third parties. RealDID's open-wallet policy (§5.6) admits a plurality of approved wallets and additionally allows issuers to host their own wallet for credentials they issue, removing third-party wallet operators from the issuance and use loop entirely in that configuration. The privacy-relevant question therefore shifts from "is the wallet user-controlled?" (both architectures answer yes) to "how many wallet operators sit between holder and verifier, and what do they observe?" — and on this dimension RealDID's open-wallet policy yields a smaller and more configurable observation surface than the single-wallet EUDI model.

### *6.1.2 GDPR, PIPL, and Erasure Against an Immutable Ledger*

A serious comparative-law gap that the centralized-versus-decentralized framing tends to glide past is the tension between *immutable* on-chain registries and *erasure* rights under modern data-protection regimes. We analyze this for the two regimes most relevant to RealDID's cross-border scope: China's PIPL [6] and the European Union's GDPR [17].

PIPL Art. 47 grants data subjects a right to deletion under specified conditions; GDPR Art. 17 grants a broader right to erasure ("right to be forgotten") with explicit exceptions. Recent comparative analyses note that PIPL is structurally closer to GDPR than to U.S. sectoral regimes on data-subject rights, making the GDPR contrast a tight rather than distant comparison for our purposes [28, 29]. Both rights presuppose that a controller can, in principle, delete personal data on request. A public, permissioned, append-only ledger that records DIDs, template registrations, and revocation entries cannot satisfy this presupposition mechanically: even a "deactivated" DID remains a permanent on-chain record, and historical entries cannot be redacted without breaking the ledger's integrity.

RealDID's design mitigates but does not resolve this tension along three axes. *First*, the on-chain record contains only the DID and a public key; it does not contain directly identifying personal data — the linkage to CTID is held off-chain by the state. To the extent that a DID alone qualifies as personal data (it is plausibly pseudonymous data under GDPR Recital 26 and likely "personal information" under PIPL), the on-chain residue after revocation is bounded but non-zero. *Second*, revocation entries published on chain mark a DID as inactive but cannot remove it. This is closer to GDPR's notion of *restriction of processing* (Art. 18) than to outright erasure, and it requires policy alignment that PIPL and GDPR jurisprudence have not yet fully developed for ledger-based identity infrastructure. *Third*, redactable-blockchain constructions [18] offer a theoretical path to ledger-level erasure but are not part of the current RealDID design and carry their own governance complications (who holds the redaction key?).

We do not claim that RealDID satisfies GDPR Art. 17 today. We claim, more modestly, that the architecture's choice to keep all directly identifying data off-chain narrows the surface of conflict between immutability and erasure rights to *pseudonymous* identifiers and *registry* entries, and that the residual conflict is real and requires either policy reform, ledger-level redaction, or short-lived/rotating DIDs as part of future iterations.

## 6.2 The Dual-Root Trust Model and Its Asymmetry

The RealDID architecture establishes two roots of trust that constrain each other along distinct dimensions, but the constraints are not symmetric (Table 2, §4.4).

**State Root (CTID to RealDID).** The state guarantees that every `did:sic:` maps to a real, verified legal person. This is the *anchoring function*: without it, DIDs are self-asserted and VCs built on them carry no legal weight. The state root is hard-bounded in three ways: the state cannot sign as any DID subject (no private key access), cannot view VC contents (encrypted to the holder's public key), and cannot unilaterally modify the DID Registry, Template Registry, or Revocation Registry (multi-stakeholder consortium governance). It nevertheless retains three structural powers — it can deanonymize a DID via the linkage table (under defined legal process), it can revoke a DID and thereby cascade-invalidate all associated VCs, and it observes per-credential metadata at the relay (issuer, holder, template, timestamp, ciphertext hash).

**User Root (Private Key).** The user guarantees that only they can present their credentials and apply for new VCs. This is the *sovereignty function*: without it, the state could impersonate any citizen in digital transactions. The user root is hard-bounded as well: the user cannot obtain a DID without CTID verification, cannot fabricate a legal identity, cannot prevent deanonymization under lawful process, and cannot prevent state-side observation of credential-graph metadata.

**Asymmetric, not mutual.** It would be tempting to call this a *mutual* constraint, and earlier drafts of this work did. We now resist that framing for the reasons established in §4.4: the user is bounded by cryptography; the state is bounded by cryptography plus legal-process commitments around capabilities it both holds and operates — categorically different bounds. The architecture is therefore a departure from both pure self-sovereign identity, which assumes the user can bootstrap their own legal existence, and pure state-issued digital identity, which assumes the state can be trusted with unilateral control of all identity functions. The trade-off it offers is: stronger privacy and verifier-side decentralization, at the cost of accepting a structurally privileged state root.

## 6.3 Accountability Without Mass Surveillance

A persistent tension in digital identity architecture is the apparent trade-off between accountability and privacy. The RealDID architecture suggests this trade-off is not inevitable.

**Anonymous to Verifiers.** Verifiers see DIDs – opaque strings – and the specific claims the holder chooses to disclose. A verifier can confirm that a credential is authentic, legally anchored, and presented by its owner, without learning the holder's name, national ID number, photograph, or any other PII. This is true for all verifiers, domestic and foreign.

**Accountable to Law.** The government holds the RealDID-CTID linkage table. If a DID is used for criminal activity, the government can identify the real person behind it through a defined legal process (judicial warrant). This is not a backdoor; it is a transparent architectural property. The linkage table is not a real-time surveillance feed – it is an accountability mechanism gated by legal process.

**Audit Trail with Content Blindness.** The government logs metadata for every VC relay: issuer DID, holder DID, template ID, VC hash, and timestamp. This log supports audit and accountability without exposing credential contents. The government can answer "was a VC of this type issued to this holder at this time?" but cannot answer "what degree did this holder earn?" or "what medical condition does this VC attest to?"

**Verifier Privacy.** Unlike the centralized model, where every verification API call reveals the verifier's identity to the government, on-chain verification requires no authentication. This is the realization of **G5** (verifier privacy): the verification protocol itself creates no observation point — there is no structural equivalent of the CTID gateway that logs every query. The strength of the guarantee, however, depends on the verifier's operational choice. For verifiers operating their own read-only Guoxin Chain full node (see §4.2), the property is absolute at the application layer against both **A3** (semi-honest government relay) and **A4** (network observer): verification queries never traverse a third-party RPC endpoint and are not visible to any consortium operator, including government-agency consortium members. For verifiers using a shared consortium RPC endpoint, the query remains unauthenticated — no API key or identity token is required — but the RPC operator sees the verifier's IP address and the queried DID. The RPC operator may be a government agency (§4.2 lists government agencies among consortium members), in which case the government *can* observe query metadata at that endpoint. This is not a protocol-level observation — the verifier controls the trade-off: the architecture enables full-node privacy, an option the architecture does not foreclose, which distinguishes it from a centralized API model where verifier exposure is structurally unavoidable. A company verifying an employee's credentials, a clinic verifying a patient's insurance, a journalist verifying a source's institutional affiliation — all of these relationships are exposed to the government in a centralized verification model. RealDID protects them as a structural property of on-chain queries whose strength is under the verifier's operational control.

## 6.4 Risks and Honest Tensions

A credible policy analysis must acknowledge the architecture's risks and limitations.

**Linkage Table Compromise.** The RealDID-CTID linkage table is the single highest-value target in the system. If compromised, all RealDID holders are simultaneously de-anonymized. This is the scenario in which **G6** (accountability under legal process) degrades to mass surveillance: the gated deanonymization capability becomes an uncontrolled disclosure. In terms of the adversary model, the threat is **A3** (semi-honest government relay) exceeding its defined bounds — or an external attacker exploiting the concentration of identity data that the state-root model structurally requires. Mitigations include off-chain storage with layered encryption, hardware security modules, strict access controls, and audit logging of all access attempts. However, no mitigation can reduce this risk to zero. The linkage table concentrates risk in a way that is inherent to the state-root model.

**Credential-Graph Metadata Accumulation.** As discussed in §5.5, the government relay logs per-issuance metadata `{issuer DID, holder DID, template ID, VC hash, timestamp}`. Aggregated across the population, this constitutes a *credential graph* — for every citizen, the set of credential types they hold, the issuers they hold them from, and the issuance times. This is **A3**'s (semi-honest government relay) structural observation surface: the protocol *requires* the relay to log this metadata for audit, and the relay therefore accumulates it lawfully, but the aggregation itself constitutes a surveillance dataset whose sensitivity is independent of credential contents. It reveals affiliations (which university, which hospital, which professional body) and life events (issuance of a marriage credential, a residence permit, a clearance). The metadata-privacy literature shows that comparable graphs (call records, transaction records) often re-identify individuals and disclose far more than nominally "content-free" framing would suggest [7, 8]. The current architecture does not protect against this. Plausible mitigations include template-ID obfuscation or commitment schemes, blinded-issuance constructions [9], or strict retention and access controls on the relay log. We treat these as priority items for future iterations of the design rather than features of the system as currently deployed.

**Presentation Linkability.** The RealDID design as described here uses a single `did:sic:` per holder across all presentations. Two consequences follow. First, multiple **A1** (malicious verifiers) can collude — or one verifier across multiple sessions can self-correlate — to link a holder's interactions, since each presentation carries the same DID-bound holder signature against the same on-chain DID document. Verifier collusion is one of the specific capabilities attributed to A1 in §3.0. Second, even non-colluding verifiers who simply log presentations can produce, in aggregate, a profile of the holder's verification activity tied to a stable identifier. The architecture therefore provides *content* selective disclosure but not *unlinkability* in the cryptographic sense achieved by anonymous-credential schemes such as BBS+ signatures [10], Camenisch-Lysyanskaya signatures [11], or IDEMIX-style constructions [24], and not the disclosure-time unlinkability targeted by the IETF SD-JWT draft [12]. Mitigations available within the current model include per-relationship or per-presentation pairwise DIDs (as in the W3C DID core specification's privacy guidance [1]) and short-lived DIDs reissued by the state on demand; both carry usability and operational costs. We mark this as a structural limitation rather than a flaw — the architecture's primary goal was legal anchoring rather than full unlinkable presentation — but the limit deserves to be named explicitly and is not adequately addressed by content blindness alone.

**Revocation Power.** The government can revoke a RealDID, which cascade-invalidates all VCs linked to that DID. This is the flip side of **G6** (accountability under legal process): the same capability that enables lawful deanonymization also enables a single action to invalidate all of a person's credentials. It is necessary for addressing identity fraud, criminal misuse, or national security concerns. However, the breadth of the power warrants scrutiny. What are the checks on revocation? Is there a process for appeal? Can revocation be scoped (e.g., revoke only newly-issued VCs, preserving existing ones)? These are governance questions that the architecture itself does not answer.

**Template Approval Gatekeeping.** The government approves all VC templates before they are registered on-chain. This ensures that only authorized issuers can create credential types within their domain of authority — the defense against **A2** (semi-honest issuer) attempting to issue credentials outside its legitimate scope. However, it also creates a gatekeeping point: the government could theoretically block a template for reasons unrelated to issuer authorization — for example, a credential type that enables disfavored forms of association or commerce. The multi-stakeholder governance of the Template Registry provides some check on this power (no single entity can unilaterally approve or reject), but the risk of gatekeeping creep is real.

**State Power and the Permissioned Chain.** The open permissioned chain is governed by a consortium that includes government agencies. This is the institutional context within which **A5** (consortium-majority adversary) must be evaluated: the defense against A5 relies on consortium consensus rather than cryptographic enforcement. While the consortium structure prevents unilateral action, a colluding majority of consortium members — or a state that exercises effective control over enough consortium members — could theoretically alter the chain's state. On a public permissionless chain, collusion requires economic majority (51% attack); on a permissioned consortium chain, it requires institutional capture — a different threat model with a different trust assumption. The paper does not claim equivalence; it claims that open permissioned chains achieve the minimum necessary properties for decentralized verification under China's legal constraints.

## 7. Generalizability

A case study of a single system risks being dismissed as a point solution. This section demonstrates that the RealDID three-layer model generalizes to any country with a centralized legal identity system, and that the model is chain-agnostic – deployable on open permissioned chains where cryptocurrency is restricted, and on public permissionless chains where legally permissible.

### 7.1 The Three-Layer Framework

The RealDID architecture abstracts to a three-layer model that is independent of any specific national implementation:

1. **Layer 1 – National Legal Identity.** A centralized system that verifies legal identity through biometric and documentary evidence. Examples: China's CTID, India's Aadhaar, Estonia's eID, and the national identity systems of Singapore and Hong Kong.
2. **Layer 2 – Decentralized Identifier (DID).** A DID issued by the state onto a verifiable data registry (blockchain), with user-generated key pairs, state-held linkage, and public resolvability. The DID serves as the decentralized anchor for credential exchange.
3. **Layer 3 – Verifiable Credentials.** A rich ecosystem of issuer-defined VC templates, government-mediated content-blind issuance, triple-signature authentication, and selective disclosure at presentation.

The chain is a deployment parameter of Layer 2, not an architectural requirement of the model. The model prescribes *properties* – public readability, DID resolvability, cryptographic verifiability, multi-stakeholder governance of registries – not a specific chain type.

### 7.2 Chain Agnosticism

The three-layer model deploys differently depending on the legal status of public permissionless blockchains in a given jurisdiction.

**Where public permissionless chains are legal** (e.g., Singapore, Hong Kong), the architecture deploys directly on an existing public chain (Ethereum, Polygon, etc.). Chain consensus is already decentralized – the governance challenge shifts from the chain layer to the smart contract layer. The DID Registry, Template Registry, and Revocation Registry are implemented as smart contracts governed by multi-stakeholder mechanisms (multi-signature wallets, DAO-style governance, or institutional consortium agreements). The state issues DIDs as transactions to these contracts (requiring gas fees, which are permissible in these jurisdictions); verifiers query the public chain without restriction.

**Where public permissionless chains are restricted** (e.g., China), the architecture deploys on an open permissioned chain with consortium governance. The same three smart contracts exist, but the validator set is permissioned to consortium members, and the chain itself requires consortium governance.

In both cases, the essential guarantee is identical: **no single entity — including the state — can unilaterally alter DIDs, approve templates, or revoke credentials**. The mechanism differs (smart-contract governance on public chains; chain-plus-contract governance on permissioned chains), but the architectural property is preserved. Table 4 summarises the differences across jurisdictions.

| | China | Singapore | Hong Kong |
|---|---|---|---|
| **Layer 1** | CTID | National ID system | National ID system |
| **Chain type** | Open permissioned | Public permissionless | Public permissionless |
| **Constraint** | Crypto gas fees illegal | No legal restriction | No legal restriction |
| **Governance target** | Chain + smart contracts | Smart contracts only | Smart contracts only |

*Table 4: Chain-agnostic deployment of the three-layer model across jurisdictions.*

The generalizability of the model is not theoretical. Cross-border deployments are in progress with Singapore and Hong Kong, testing the three-layer model in practice: DIDs anchored to each jurisdiction's national identity system, cross-border VC exchange without exposing identity infrastructure across borders, and the Chinese permissioned chain serving as a neutral verification ground. Results will be reported in future work.

## 7.3 Prerequisites for Adoption

Deploying the three-layer model in a new jurisdiction requires five prerequisites, summarized in Table 5.

| Prerequisite | Rationale |
|---|---|

| Prerequisite | Rationale |
|---|---|
| Centralized legal identity with biometric verification | Provides the Layer 1 root of trust – DIDs must anchor to verified legal persons |
| Chain with public read access and DID resolution | Verifiers anywhere must resolve DIDs and validate signatures without authorization |
| Multi-stakeholder governance of smart contracts | Prevents unilateral control of DID issuance, template approval, and revocation |
| VC template standardization | Enables cross-issuer and cross-border interoperability of credential types |
| Regulatory framework for the linkage table | Defines who may deanonymize, under what process, with what oversight |

*Table 5: Prerequisites for adopting the three-layer model.*

## 7.4 Limitations and Stakeholder Gaps

The three-layer model is not universally applicable. **Countries without centralized identity infrastructure** lack the Layer 1 root of trust and cannot adopt the model without first building a national ID system — a multi-decade infrastructure project that the model does not address. **Regions with political resistance to state involvement in digital identity** may prefer pure self-sovereign identity models in the tradition of Allen [19], accepting the absence of legal anchoring as a trade-off for freedom from state oversight; the substantial critical literature on Aadhaar's centralized identity architecture [20, 21] is a reminder that "national identity infrastructure" is not a neutral premise. **Cross-border legal recognition** remains an open challenge: a VC valid under China's PIPL may not automatically satisfy GDPR adequacy requirements (§6.1.2), requiring bilateral or multilateral legal frameworks that do not yet exist.

Beyond model-level limits, the architecture as described leaves three categories of stakeholders under-served, and we name them honestly here.

**People without devices or biometrics.** The protocol assumes a user device that generates and stores a private key (§4.3 step 2) and a population that can complete CTID biometric and documentary verification. Elderly users without smartphones, children below the biometric-enrollment age, people whose biometrics fail to enroll reliably (a known issue in large-scale biometric systems [20]), and persons with disabilities all sit outside the architecture's assumed user model. An inclusive deployment requires assisted-issuance, guardian-mediated, or delegated-presentation flows; none are specified in the current design.

**Non-citizens and persons of irregular status.** CTID is structured around verified citizens of China. Foreign residents, dual nationals, stateless persons, and irregular-status migrants either fall outside the system or require parallel infrastructure. The cross-border generalization argument in §7.1–§7.2 addresses *interoperation* between national identity roots (§7.1–§7.2) but does not address persons who have no national identity root at all.

**Key loss and recovery.** Catastrophic loss of the private key (lost device, hardware failure, key compromise) currently cascades to loss of every VC bound to the affected DID. Plausible recovery routes — social recovery, government-mediated reissuance through a fresh CTID re-verification, threshold-secret recovery with consortium custodians — each have distinct constraints on the trust model. Social recovery weakens the user-side cryptographic bound; government-mediated reissuance strengthens the state's structural power further. The architecture as currently specified does not commit to a recovery design; we treat this as one of the most important open questions for production deployment.

These limitations bound the model's applicability and do not undermine its value for the substantial set of countries that already possess centralized legal identity infrastructure and are seeking to bridge it to decentralized verification — but they should bound the claims made on its behalf.

## 8. Conclusion

This paper has presented China's RealDID system as a case study in bridging centralized national legal identity with decentralized verifiable credentials. The core finding is that these two paradigms are not contradictory: they can be combined into an asymmetric, state-bounded trust model — *decentralized verification with centralized legal anchoring* (§4.4, §6.2) — in which the state prevents identity fabrication but cannot impersonate or read credential contents, and the user controls presentation privacy but cannot fabricate a legal identity or evade state-side metadata observation. The broader implication is that decentralized verification can exist without decentralized identity roots, and this framing may be more practically achievable and legally compatible than the full self-sovereign identity vision, particularly in jurisdictions where the state has a legitimate and legally mandated role in establishing legal identity.

We have made three specific contributions. First, we described the RealDID three-layer architecture (CTID, RealDID, VCs) in protocol-level detail, including the DID issuance protocol, the key custody model that establishes the asymmetric trust structure, the VC lifecycle with issuer-driven template registry, the triple-signature chain (issuer, government, holder), and the SD-JWT-based selective disclosure mechanism — together with explicit acknowledgment of the architecture's metadata-level privacy limits. Second, we analyzed the architecture's policy implications: alignment with China's PIPL data minimization principle, a comparison with the EU's GDPR (including the tension between user erasure rights and immutable on-chain registries) and eIDAS 2.0, the dual-root trust model evaluated honestly as asymmetric, verifier privacy as an under-appreciated dimension of the identity ecosystem, and risks including linkage-table compromise, revocation breadth, and credential-graph metadata accumulation. Third, we argued that the three-layer model is chain-agnostic and generalizes to jurisdictions with a centralized legal identity system, supported in principle by in-progress cross-border deployments with Singapore and Hong Kong; empirical evaluation is reserved for future work.

Future work should address several open questions. Formal verification of the linkage secrecy property – proving that an observer of on-chain data cannot associate DIDs with real-world identities under defined threat models – would strengthen the privacy guarantees. Zero-knowledge proofs for selective disclosure could further reduce what verifiers learn about holders (e.g., proving age over 18 without revealing the exact birth date, even to the verifier). Cross-border legal recognition frameworks are needed to establish mutual recognition of VCs across jurisdictions with different data protection regimes. The Singapore and Hong Kong deployment results should be reported as empirical evidence for the model's generalizability. Finally, the governance questions raised in Section 6.4 – checks on revocation power, template approval oversight, and consortium capture risk – deserve dedicated treatment as the model moves from sandbox to production.

## Acknowledgements

In preparing this manuscript, the author used Anthropic's Claude as an AI-assisted tool. The assistance includes literature search, copyediting, grammar, and proofreading, and is therefore disclosed here in full, in accordance with the publisher's policy on generative AI in scientific writing. All AI-assisted output was reviewed, edited, and approved by the author.

## Competing Interests

The author, Yifan He, is the Chief Executive Officer of Red Date Technology (Hong Kong) Ltd. Red Date Technology is the architect and principal operator of China's Blockchain-based Service Network (BSN) and is, together with entities under China's State Information Center, a co-developer of the China RealDID system that is the subject of this paper. Red Date Technology also operates Guoxin Chain, the open permissioned chain on which RealDID is deployed, and is a stakeholder in the Singapore and Hong Kong cross-border sandbox engagements discussed in §7.2.

The author therefore has a direct material and professional interest in the favorable reception of the architecture analyzed in this paper. Readers are asked to weigh the paper's normative and comparative claims (in particular: claims of regulatory alignment in §6.1, the dual-root trust model in §6.2, and the generalizability argument in §7) against this interest. The paper's descriptions of architecture and protocols (§§4–5) reflect insider knowledge and are, to the author's best understanding, accurate; readers seeking independent verification should consult the cited primary materials and the public RealDID and BSN documentation.

No external funding was received for the preparation of this paper.